\documentclass[12pt,preprint]{aastex}

\newcommand{\arcm}{\ifmmode{'}\else$'$\fi}
\newcommand{\arcs}{\ifmmode{''}\else$''$\fi}
\newcommand{\msun}{M_{\odot}}

\shorttitle{First confirmed microlens in a globular cluster}
\shortauthors{Pietrukowicz et al.}

\begin{document}

\title{The first confirmed microlens in a globular cluster\altaffilmark{\ast}}

\author{P.~Pietrukowicz\altaffilmark{1}, D.~Minniti\altaffilmark{2,3,4},
Ph.~Jetzer\altaffilmark{5}, J.~Alonso-Garc\'ia\altaffilmark{2},
A.~Udalski\altaffilmark{1}}

\begin{abstract}
In 2000 July/August a microlensing event occurred at a distance of
2\farcm33 from the center of the globular cluster M22 (NGC6656),
observed against the dense stellar field of the Milky Way bulge.
We have used the adaptive optics system NACO at the ESO Very Large
Telescope to resolve the two objects that participated in the event:
the lens and the source. The position of the objects measured in 2011
July is in agreement with the observed relative proper motion
of M22 with respect to the background bulge stars. Based on the brightness
of the microlens components we find that the source is a solar-type
star located at a distance of $6.0\pm1.5$~kpc in the bulge,
while the lens is a $0.18\pm0.01\msun$ dwarf member of the globular
cluster located at the known distance of $3.2\pm0.2$~kpc from the Sun.
\end{abstract}


\altaffiltext{*}{Based on observations collected with the ESO VLT and VISTA
telescopes at Paranal Observatory (ESO Programmes 087.C-0640(A) and
179.B-2002(B), respectively), and the 1.3-m Warsaw telescope at the Las
Campanas Observatory of the Carnegie Institution for Science.}
\altaffiltext{1}{Warsaw University Observatory,
Al. Ujazdowskie 4, 00-478 Warszawa, Poland}
\altaffiltext{2}{Departamento de Astronom\'ia y Astrof\'isica,
Pontificia Universidad Cat\'olica de Chile, Av. Vicu\~na MacKenna 4860,
Casilla 306, Santiago 22, Chile}
\altaffiltext{3}{Vatican Observatory, Vatican City State V-00120, Italy}
\altaffiltext{4}{Department of Astrophysical Sciences, Princeton University, Princeton NJ 08544-1001, USA}
\altaffiltext{5}{Institut f\"ur Theoretische Physik, Universit\"at Z\"urich,
Winterthurerstrasse 190, CH-8057 Z\"urich, Switzerland}

%%%%%%%%%%%%%%%%%%%%%%%%%%%%%%%%%%%%%%%%%%%%%%%%%%%%%%%%%%%%%%%%%%%%%%%%%%%%%%%

\section{Introduction}

The effect of gravitational microlensing of background stars by compact objects
located in globular clusters was analyzed for the first time by 
\cite{1994AcA....44..235P}. He showed that thanks to the usually well-known
distances to the source and the lens, and transverse velocity between the
populations to which the objects belong, it is possible to derive the lens
mass when the event time scale is measured. 
Paczy\'nski suggested to monitor globular clusters like M22 or 47~Tuc
in front of the rich background of either the Galactic bulge or the Small
Magellanic Cloud. According to his calculations one should detect up
to a few microlensing events in one year of continuous monitoring
of M22 with a 1~m class ground-based telescope. Some microlensing events
detected so far toward the bulge in microlensing surveys such as OGLE
\citep{2000AcA....50....1U}, MACHO \citep{2000ApJ...542..281A}, and MOA
\citep{2001MNRAS.327..868B} might be associated with globular clusters
\citep{1998A&A...336..411J,2008IJMPD..17.2305D}.

\cite{2005AcA....55..261P} presented the results of a search for erupting
objects in the field of the globular cluster M22. The cluster was observed
over 10 weeks in 2000-2001 with the 1.0~m Swope telescope at Las Campanas
Observatory as one of the targets of the Cluster AgeS Experiment
\citep[CASE;][]{2005AIPC..752...70K}. Besides two erupting
dwarf novae they found a probable microlensing event located
at $\alpha_{2000.0}=$18:36:22.40, $\delta_{2000.0}=-$23:56:29.4, i.e.,
only 2\farcm33 $=1.75~r_{\rm c}=0.69~r_{\rm h}$ from the cluster center,
where $r_{\rm c}=1\farcm33$ and $r_{\rm h}=3\farcm36$ are the core radius
and half-mass radius of the cluster, respectively (taken from the
2010 version of \cite{1996AJ....112.1487H} catalog).
The brightness of the object increased by about 0.8~mag in $V$ over 20~days.
Around 2000 August~5 it reached a maximum of $V=19.1$~mag and then
faded to a constant level of $V=19.9$~mag. Based on its color at maximum
brightness the authors excluded the possibility that the object could
be a dwarf nova. They fitted a single lens model to the light curve and
found that the most likely geometry of the event places the source
in the Galactic bulge and the lens in the cluster. The fitted parameters
are: the epoch of maximum $t_0=2451759.70^{+0.33}_{-0.34}$,
characteristic (Einstein) time $t_{\rm E}=15.9\pm1.1$~days,
impact parameter $u_0=0.54^{+0.02}_{-0.18}$ in units of Einstein
radius $r_{\rm E}$, $V_{\rm S}=19.92^{+0.62}_{-0.02}$~mag,
and $V_{\rm L}=24.8^{+\infty}_{-4.0}$~mag,
where $V_{\rm S}$ and $V_{\rm L}$ are $V$-band magnitudes
of the source and lens, respectively. The authors assessed the mass
of the lens to $M=0.14^{+0.10}_{-0.02}\msun$. Large uncertainties
of the above values result from the faintness of the object
and partial coverage of the event.

After many years, in some special cases it is possible to directly detect
the lens, measuring its mass and the geometry of the microlensing event 
\citep[e.g.,][]{2001Nature...414..617,2007ApJ...671..420K}.
In this Letter we resolve the microlensing system components based
on new near-IR high-resolution images, measuring the complete geometry
of the event, and the parameters of the source and lens stars.
The event reported here is the first confirmed
microlensing event in a globular cluster. We note that brightening
episodes detected in {\it Hubble Space Telescope} ({\it HST}) images of M22
by \cite{2001Natur.411.1022S} were later reexamined and interpreted
either as a dwarf nova type outburst \citep{2003ApJ...597L.137A}
or as a result of cosmic ray double hits \citep{2002ApJ...565L..21S}.

%%%%%%%%%%%%%%%%%%%%%%%%%%%%%%%%%%%%%%%%%%%%%%%%%%%%%%%%%%%%%%%%%%%%%%%%%%%%%%%

\section{VLT Observations and reductions}

$K_s$-band observations of the M22-microlens region were obtained
at the ESO Very Large Telescope (VLT) on 2011 July 17,
i.e., 10.95 years after the maximum of the event.
Twenty single 110~s exposures were taken using NACO at UT4, composed
of the Nasmyth Adaptive Optics System (NAOS) and the High
Resolution IR Camera and Spectrometer (CONICA). The detector was
a 1026~$\times$ 1024 pixel SBRC InSb Alladin 3 array. We used the S27
camera of the scale 27.15~mas~pixel$^{-1}$ and the field of view of
28\arcs$\times$28\arcs. The telescope jittered after each exposure
according to a random pattern in an 8\arcs$\times$8\arcs box.
As the guide source for adaptive optics (AO) image correction
we used a $V=14.1$~mag star located 11\farcs36 away from the target.
The seeing during the observations ranged between 0\farcs69 and
0\farcs99. The data were reduced with the ESO software packages
MIDAS and Eclipse. In the top panel of Figure~1 we show the combined
image of the observed field. The image is affected by anisoplanatism,
which degrades the point spread function (PSF) making it more elongated
with increasing angular distance from the guide star.
The measured full width at half maximum (FWHM) at the center
of the image is 0\farcs11. The M22-microlens region was also observed
with VLT/NACO through the $J$ filter on 2011 April 26. Unfortunately,
the measured FWHM of 0\farcs36 is insufficient to detect the faint lens.

For our analysis we cut a smaller area of 600$\times$600 pixels,
covering 16\farcs3$\times$16\farcs3 around the target microlens. We used
DAOPHOT/ALLSTAR \citep{1987PASP...99..191S} to extract photometry
of stars in the image. Due to relatively large difference in brightness
($\Delta K_s=3.2$~mag) and very small separation between the source and the
lens ($4.59~{\rm pixels}=124.6$~mas) the photometry was extracted in three steps.
In the first step we found PSF based on selected isolated bright stars.
Then using this PSF we found centroids of all stars with S/N$>3.5$.
In the second step, we removed the bright stars from the image
and extracted profile photometry for residual objects, including
the lens. The residual image showing the lens located slightly off
the center is presented in the lower panel of Figure~1.
In the final step, we re-extracted the photometry for all stars
including the positions of both the source and lens.

We performed simulations in which we inserted the same pair
of stars in the same location of 100 frames with subtracted stars in order to
assess the errors of the positions of the two objects.
The obtained mean uncertainties in pixels we converted into the unit of mas.

Standard $K_s$-band magnitudes of the stars within our field were calculated
based on photometry of 51 neighboring stars detected in the near-IR VISTA
Variables in the Via Lactea survey \citep[VVV;][]{2010NewA...15..433M}.
We found the source and lens to have $K_s=17.37\pm0.03$~mag and
$K_s=20.57\pm0.09$~mag, respectively.

%%%%%%%%%%%%%%%%%%%%%%%%%%%%%%%%%%%%%%%%%%%%%%%%%%%%%%%%%%%%%%%%%%%%%%%%%%%%%%%

\section{Confirmation of the microlensing event}

Almost eleven (10.95) years after the microlensing event we found
the lens located ($123.6\pm1.8$, $15.8\pm1.8$)~mas (east, south)
from the source. This corresponds to a relative proper motion of the lens
with respect to the source
$[\mu_{\alpha} {\rm cos} \delta,\mu_{\delta}]=[11.29\pm0.17,-1.44\pm0.17]$~mas~yr$^{-1}$
and its total value $\mu_{\rm rel}=11.38\pm0.24$~mas~yr$^{-1}$.
Based on archival {\it HST} observations \cite{2004ChPhL..21.1673C}
measured the proper motion of the globular cluster M22 with respect
to the background bulge stars. They obtained
$[\mu_{\alpha} {\rm cos} \delta,\mu_{\delta}]=[10.19 \pm 0.20,-3.34 \pm 0.10]$~mas~yr$^{-1}$
and showed that the separation between cluster and field stars is clear.
They considered all stars with proper motions $<2$~mas~yr$^{-1}$ around
the mean value of the cluster to be M22 members, and stars with motions
$>\mu_{\alpha} {\rm cos}\delta=5$~mas~yr$^{-1}$ as mainly bulge stars.
In a vector-point diagram presented in Figure~2 we overlaid
the vector measured here for the microlens on the vector for the bulge-M22 set
from \cite{2004ChPhL..21.1673C}. The microlens vector originates from
the (0, 0) point, which refers to the cluster, and ends well within
the bulge area. This confirms the geometry of the microlensing
event with the source in the bulge and the lens in the globular cluster.

By fitting a model to the light curve \cite{2005AcA....55..261P} predicted
that the lens is fainter than the source by $\sim5$~mag in the $V$ band.
At a distance $d_{\rm M22}=3.2$~kpc and mean reddening $E(B-V)=0.38$~mag
\citep{2004MNRAS.349.1278M} it is likely an M5 dwarf
of an absolute brightness $M_V\sim11.1$~mag.
Such star observed in the $K_s$ band would have $\sim20.5$~mag
\citep[based on models from][]{1998MNRAS.295..711B}. The brightness
of the faint object detected close to the target source in the VLT/NACO
image is $K_s=20.57\pm0.09$~mag, which is in excellent agreement.

The VLT/NACO $K_s$-band image is the only available image
containing both microlensing system components. We searched
the {\it HST} archives for other high-resolution images. Unfortunately,
in two {\it HST}/{\it Advanced Camera for Surveys} (ACS)
 images taken as a part of the GO 10775 program on 2006 Apr~1
our target lies 1\farcs3 off the edge. The only {\it HST} image 
(jb1w01010, GO 11558) covering the M22 microlens was obtained
on 2010 Mar~2 in the O\textsc{[iii]} filter centered at 5023\AA~and
with FWHM=86\AA. We checked that in this narrow-band filter all
objects of similar brightness to the lens in the NACO image are
below the detection limit. This supports the fact that
the lens is a relatively red object.

We also checked brightness variations of the target object in recent
OGLE data obtained with the 1.3~m Warsaw telescope at Las Campanas
Observatory, Chile. The Optical Gravitational Lensing Experiment
during its fourth phase (OGLE-IV), that started in 2010 March,
observes the globular cluster M22 occasionally once or twice a week.
In Figure~3 we present the $I$-band light curve of the target object
in years 2010-2011. The zero point accuracy of the magnitude scale
is about 0.1~mag. Constant brightness of the object within 0.2~mag
corroborates that the episode of increasing brightness
in 2000 July/August was a single event.

Theoretically we can estimate the probability of a chance configuration of
two unrelated stars in the investigated area. We detected 342 stars in a brightness
range $15.4<K_s<22.6$~mag in the 15\arcs$\times$15\arcs~field centered
on the target source. Assuming Poisson statistics this gives a density
of $1.52\pm0.08$ stars arcsec$^{-2}$ or $0.074\pm0.004$ stars within 124.6~mas
around the target. Eighty-two stars (corresponding to a fraction of 0.240), being fainter
than $K_s=20$~mag, could act as a potential lens in our case. The acceptable position
angle of the lens ranges within $\pm38\fdg5$ off the M22-bulge relative proper
motion direction, decreasing the chance by 0.214. If we take into account
all above requirements we find $0.38$\%$\pm0.02$\% chance of such configuration
at any location in the field. However, the observed position and brightness
of both lens and source being in perfect agreement with the expectations unambiguously
confirm the microlensing nature and geometry of the event detected in 2000.

%%%%%%%%%%%%%%%%%%%%%%%%%%%%%%%%%%%%%%%%%%%%%%%%%%%%%%%%%%%%%%%%%%%%%%%%%%%%%%%

\section{Masses and distances to the microlens components}

The observed $K_s$-band brightness of the lensing star and the fact
that it is located in the globular cluster M22 allows us to determine
its type. According to \cite{2004MNRAS.349.1278M} M22 lies at a
distance $d_{\rm M22}=3.2\pm0.2$~kpc from the Sun and has an average
metallicity [Fe/H]$_{\rm CG}=-1.68\pm0.15$~dex in \cite{1997A&AS..121...95C}
scale. Reddening in the direction of M22
is spread between $E(B-V)=0.34$ and 0.42~mag \citep{1999A&A...350..476R}.
Using \cite{1985ApJ...288..618R} relations on absorption, that $A_K=0.112 A_V$,
where $A_V=3.1 E(B-V)$, we find $0.118<A_K<0.146$~mag for objects in M22.
From this we obtain the absolute brightness of the lens $M_{Ks}=7.91\pm0.16$~mag.
Based on models from \cite{1998MNRAS.295..711B} we find the mass of the star
$M_{\rm lens}=0.18\pm0.01\msun$ (see Figure~4).

Knowing the distance to the lensing object $d_{\rm lens}$,
its mass $M_{\rm lens}$, relative proper motion $\mu_{\rm rel}$
between the source and lens, and time scale of the event $t_{\rm E}$
we can estimate distance to the source from the following relation
$$
d_{\rm source}=d_{\rm lens}\Bigg(1-\frac{c^2\mu_{\rm rel}^2t_{\rm E}^2d_{\rm lens}}{4GM_{\rm lens}}\Bigg)^{-1},
$$
where $G$ is the gravity constant and $c$ the speed of light.
The quantities $\mu_{\rm rel}$ and $t_{\rm E}$ should
be given in either heliocentric or geocentric frame.
For the microlens in M22 we obtain $d_{\rm source}=6.0\pm1.5$~kpc which
places the source in the Galactic bulge, as expected from the relative motion.
The large errors reflect mainly the uncertainty in the estimated duration of
the microlensing event.

According to \cite{1998ApJ...500..525S} the total reddening in the cluster direction
amounts to $E(B-V)=0.33$~mag. That implies that any stars located in the cluster
field cannot be significantly more reddened than the cluster itself.
If we assume the same absorption for the source located at 6.0~kpc as for the
cluster, $A_K=0.13$~mag, from the observed brightness of the source $K_s=17.37$~mag
we find it to be a solar-type star \citep{2006ApJ...642..797P}.
Location of this object in a $K_s$ versus O\textsc{[iii]}$-K_s$ diagram
shown in Figure~5 supports this conclusion.

%%%%%%%%%%%%%%%%%%%%%%%%%%%%%%%%%%%%%%%%%%%%%%%%%%%%%%%%%%%%%%%%%%%%%%%%%%%%%%%

\section{Summary}

In the Letter we have shown that the microlensing event which occurred $2\farcm33$
from the center of the globular cluster M22 in 2000 July/August involved a
$0.18\pm0.01\msun$ dwarf of the cluster and a background solar-like star
located in the Galactic bulge. Almost 11 years after the event, using
high-resolution near-IR image we resolved the two microlensing components.
The observed position of the source and lens stars
as well as their brightness are consistent with the proposed
earlier geometry of the event. Additional evidence comes from the constant
brightness of the target object in the last two years (2010-2011).

%%%%%%%%%%%%%%%%%%%%%%%%%%%%%%%%%%%%%%%%%%%%%%%%%%%%%%%%%%%%%%%%%%%%%%%%%%%%%%%

\acknowledgements

We thank M. Jaroszy\'nski for helpful discussions, and A. Gould for
drawing our attention to an inconsistency in our original calculation
of the source distance. P.P. and A.U. are supported by funding
to the OGLE project from the European Research Council under
the European Community's Seventh Framework Programme
(FP7/2007-2013)/ERC grant agreement no. 246678. P.P. is also supported
by the grant no. IP2010 031570 financed by the Polish Ministry
of Sciences and Higher Education under Iuventus Plus programme.
We gratefully acknowledge use of data from the ESO Public Survey
programme ID 179.B-2002 taken with the VISTA telescope, and data products
from the Cambridge Astronomical Survey Unit. D.M. and J.A.-G. are supported
by Proyecto Fondecyt Regular 1090213, the BASAL Center for Astrophysics
and Associated Technologies PFB-06, the FONDAP Center for Astrophysics 15010003,
and the Milky Way Millennium Nucleus from the Ministry of Economia ICM
grant P07-021-F.

%%%%%%%%%%%%%%%%%%%%%%%%%%%%%%%%%%%%%%%%%%%%%%%%%%%%%%%%%%%%%%%%%%%%%%%%%%%%%%%

\begin{figure}
\centering
\vspace{-0.8 cm}
\includegraphics[width=6.0cm]{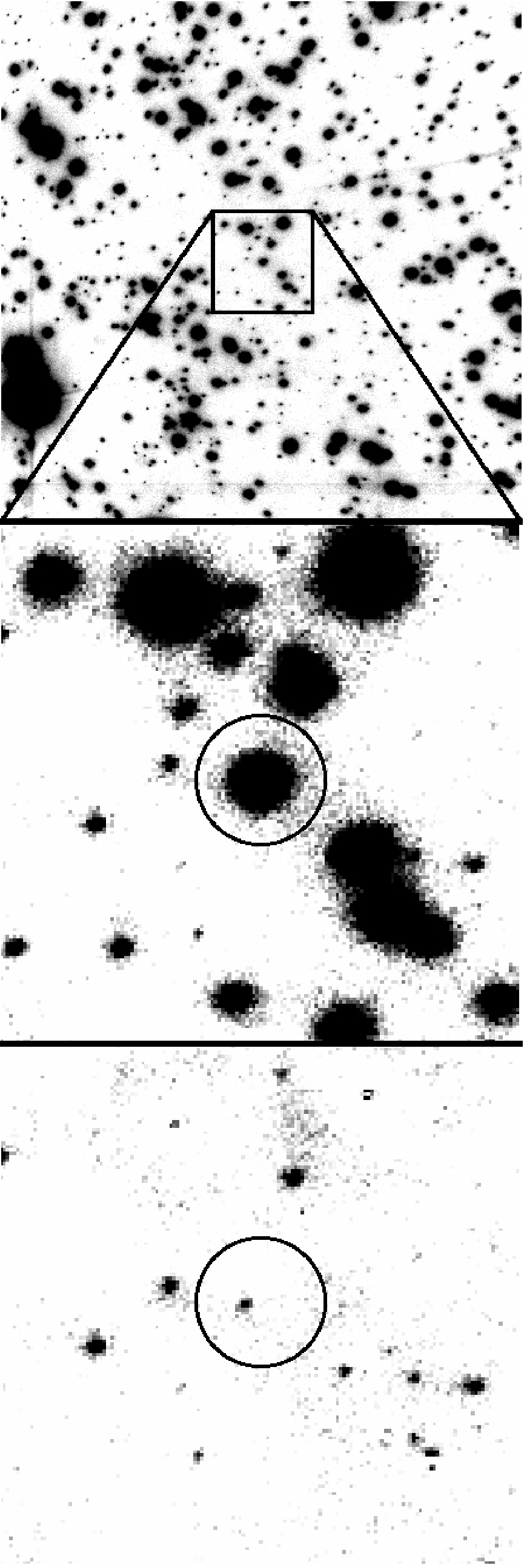}
\caption{$K_s$-band images of the microlens in M22. The field of view
in the top panel is 20\arcs on a side. North is up and east is to
the left. The brightest star near the SE corner of the top image
served as the natural guide source for the AO image correction.
The 4\arcs$\times$4\arcs close-up view centered on the target
source is presented in the middle panel. The lower panel shows a residual
image after subtracting bright stars. The faint residual object located
slightly off the center is the lensing star.
}
\label{fig:charts}
\end{figure}

\begin{figure}
\centering
\vspace{-0.8 cm}
\includegraphics[width=1.0\columnwidth]{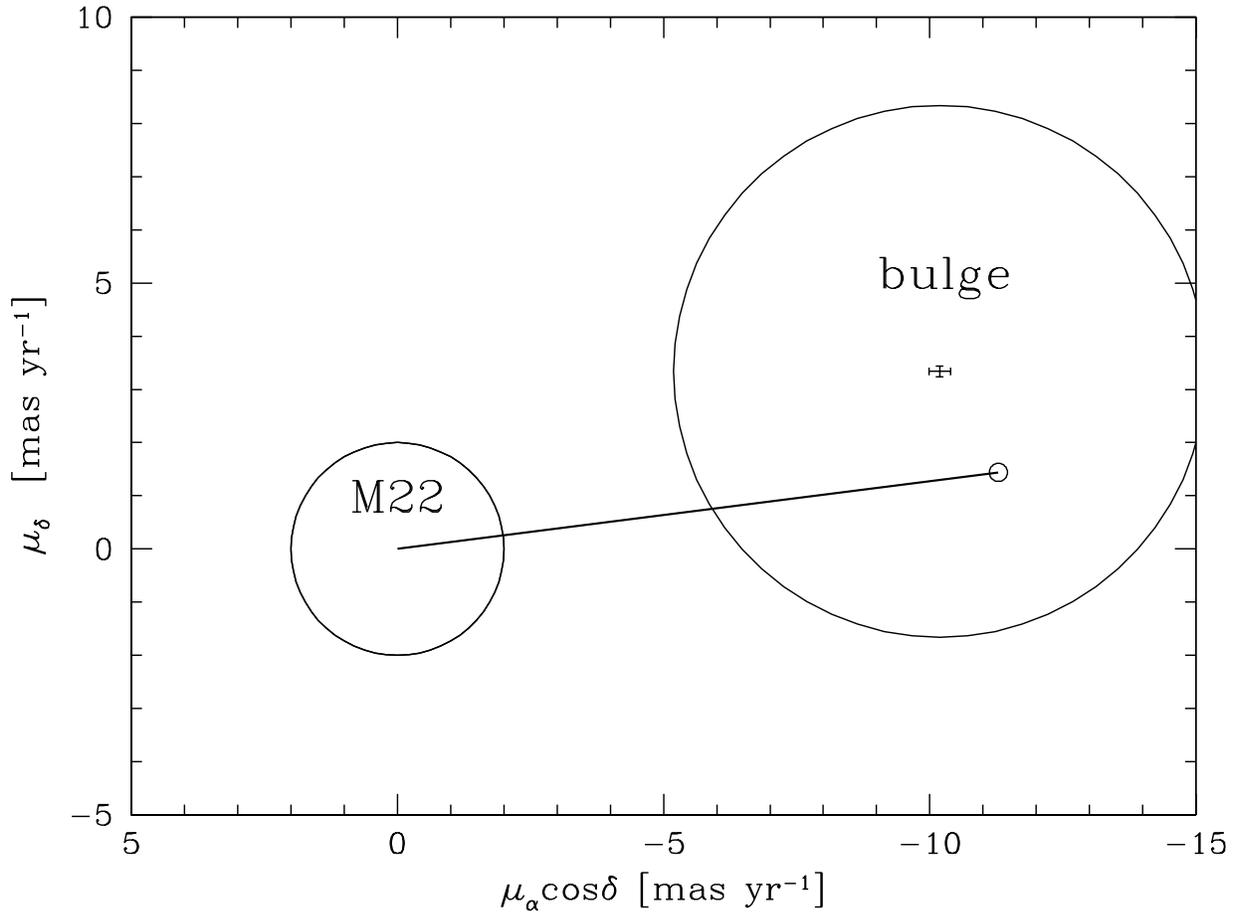}
\caption{Vector-point diagram of relative proper motions in the J2000
equatorial coordinate system of the bulge with respect to the globular
cluster M22 \citep[based on][]{2004ChPhL..21.1673C}. Stars that
would fall inside the circle of radius 2~mas~yr$^{-1}$
centered at (0, 0) are considered to be cluster members, while stars
that would fall inside the circle of radius 5~mas~yr$^{-1}$ centered at
($-10.19$, 3.34)~mas~yr$^{-1}$ are very likely bulge stars. The relative
motion between the microlens system components is shown as
the solid line with the small circle representing the uncertainty.
The length and direction of the vector unambiguously confirm that
the source belongs to the bulge, and the lens to the cluster.}
\label{fig:relmu}
\end{figure}

\begin{figure}
\centering
\vspace{-0.8 cm}
\includegraphics[width=1.0\columnwidth]{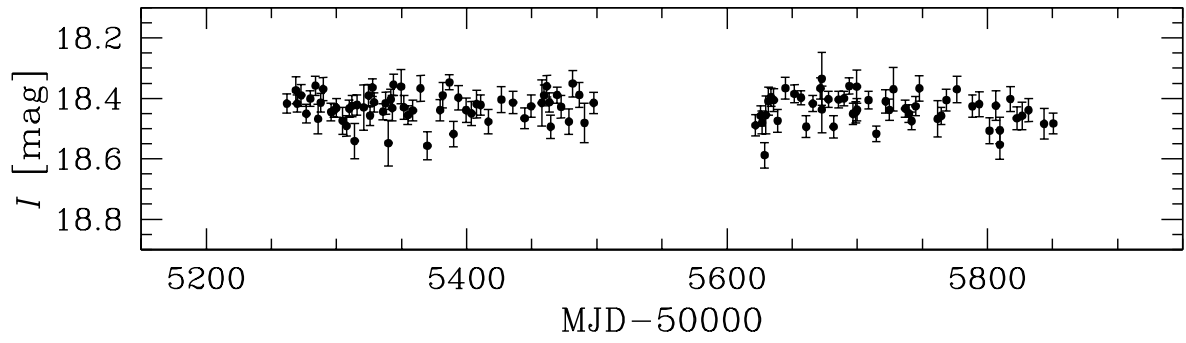}
\caption{OGLE-IV $I$-band light curve of the target object in years 2010-2011.
Its constant brightness within 0.2~mag confirms the microlesing nature
of the event in 2000 Jul/Aug.}
\label{fig:OGLElc}
\end{figure}

\begin{figure}
\centering
\vspace{-0.8 cm}
\includegraphics[width=1.0\columnwidth]{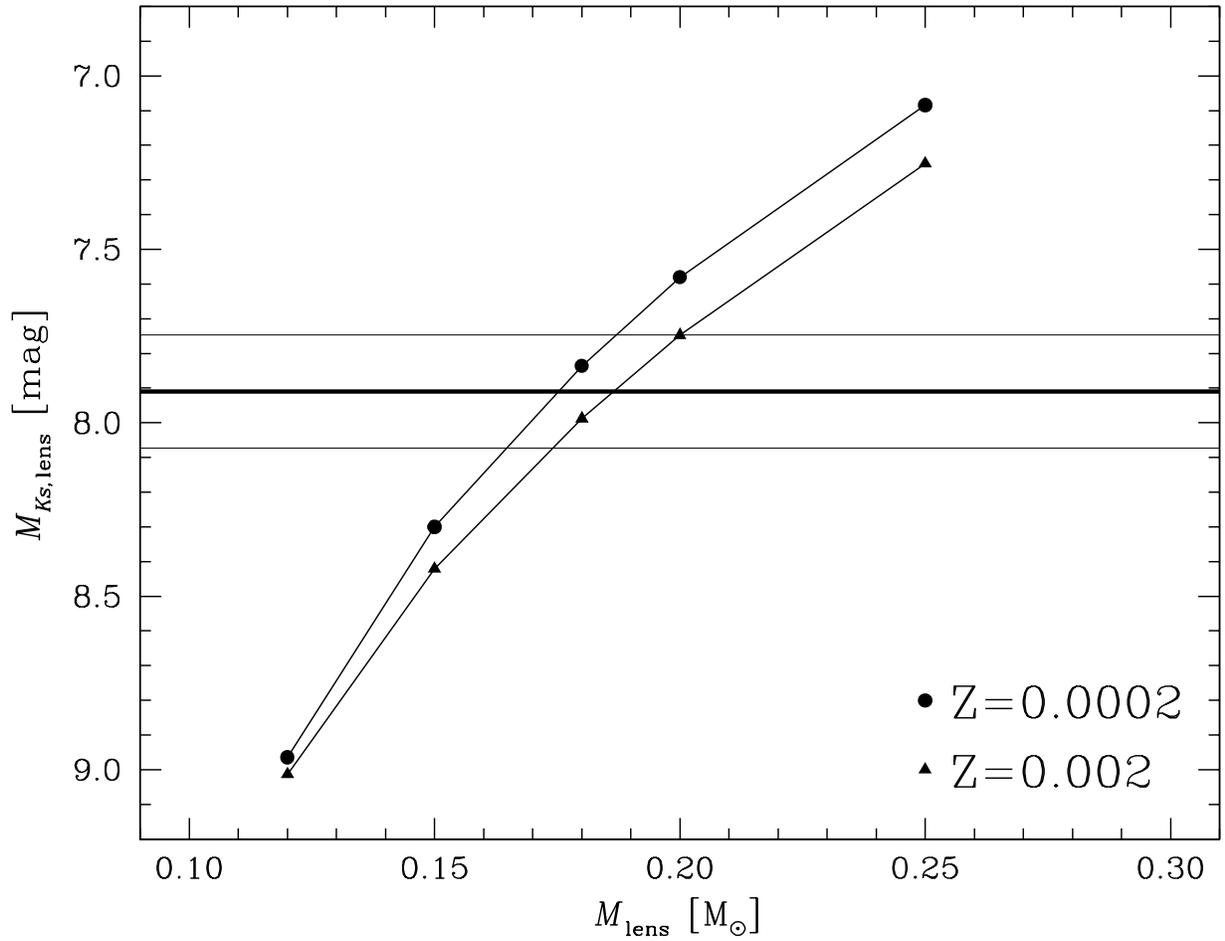}
\caption{$K_s$-band absolute brightness for dwarfs of five different masses
and two metallicities, $Z$=0.0002 and 0.002, corresponding to [Fe/H]=$-2.00$
and $-0.96$~dex, respectively \citep[data points taken from][]{1998MNRAS.295..711B}.
The measured brightness of the lens and its uncertainty are marked with
the horizontal lines.}
\label{fig:MKM}
\end{figure}

\begin{figure}
\centering
\vspace{-0.8 cm}
\includegraphics[width=12.0cm]{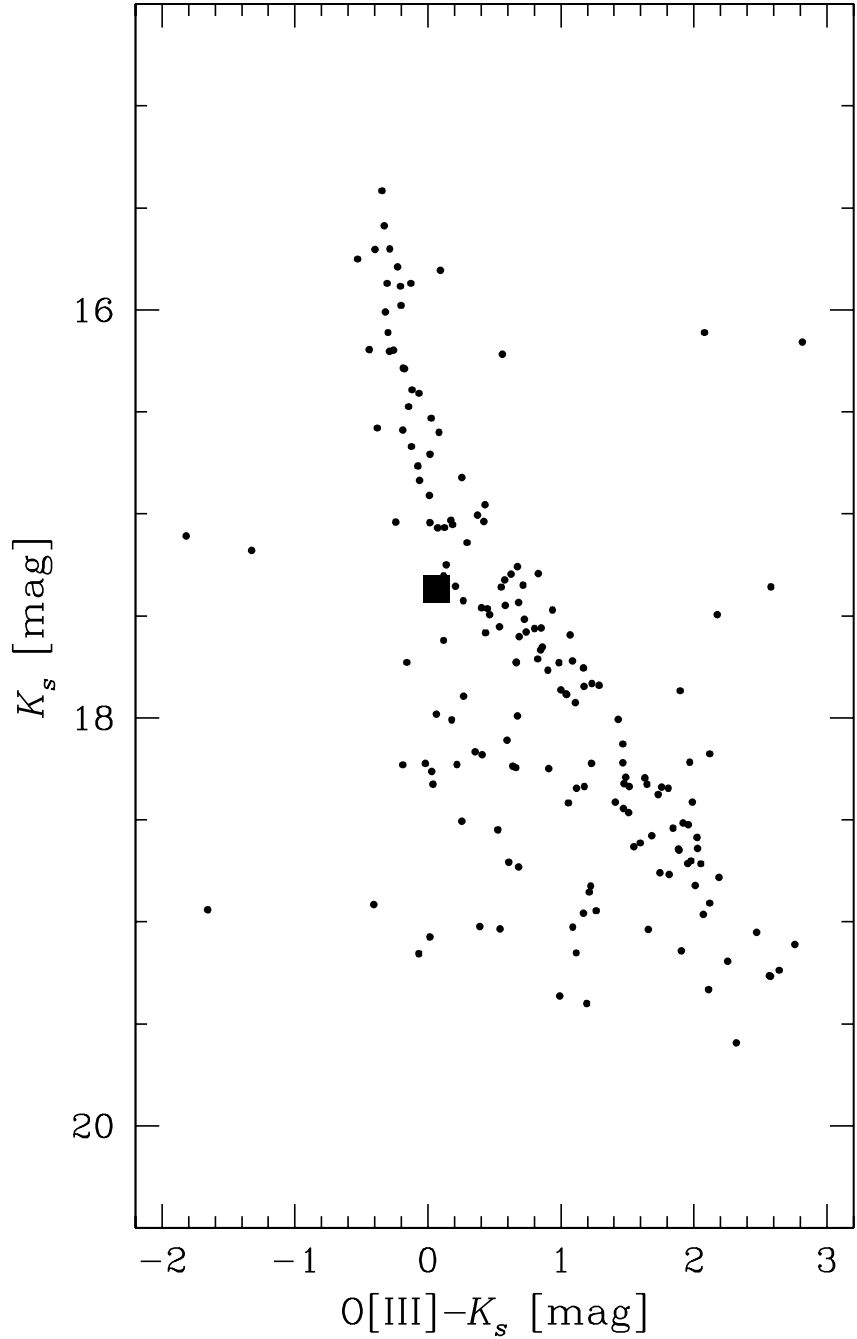}
\caption{$K_s$ vs. O\textsc{[iii]}$-K_s$ diagram for 175 stars present in
both VLT/NACO and {\it HST}/ACS images in the M22-microlens area. The majority of the
objects are main-sequence (MS) stars of the cluster. Location of the more distant
source (marked with the filled square) shows that it is fainter than the M22
MS, and consistent with a bulge MS star.}
\label{fig:cmd}
\end{figure}


\begin{thebibliography}{}

\bibitem[Alcock et al.(2000)]{2000ApJ...542..281A} Alcock, C., Allsman, R.~A., Alves, D.~R., et al. 2000, \apj, 542, 281
\bibitem[Alcock et al.(2001)]{2001Nature...414..617} Alcock, C., Allsman, R.~A., Alves, D.~R., et al. 2001, Nature, 414, 617
\bibitem[Anderson et al.(2003)]{2003ApJ...597L.137A} Anderson, J., Cool, A.~M., \& King, I.~R. 2003, \apj, 597, L137
\bibitem[Bond et al.(2001)]{2001MNRAS.327..868B} Bond, I.~A., Abe, F., Dodd, R.~J., et al. 2001, \mnras, 327, 868
\bibitem[Brocato et al.(1998)]{1998MNRAS.295..711B} Brocato, E., Cassisi, S., \& Castellani, V. 1998, \mnras, 295, 711
\bibitem[Carretta \& Gratton(1997)]{1997A&AS..121...95C} Carretta E., \& Gratton R.~G. 1997, \aaps, 121, 95
\bibitem[Chen et al.(2004)]{2004ChPhL..21.1673C} Chen, D., Chen, L., \& Wang, J.-J. 2004, Chin. Phys. Lett., 21, 1673
\bibitem[de Luca \& Jetzer(2008)]{2008IJMPD..17.2305D} de Luca, F., \& Jetzer, Ph. 2008, Int. J. Mod. Phys. D, 17, 2305
\bibitem[Harris(1996)]{1996AJ....112.1487H} Harris, W.~E. 1996, \aj, 112, 1487
\bibitem[Jetzer et al.(1998)]{1998A&A...336..411J} Jetzer, Ph., Str\"assle, M., \& Wandeler, U. 1998, \aap, 336, 411
\bibitem[Kaluzny et al.(2005)]{2005AIPC..752...70K} Kaluzny, J., Thompson, I.~B., Krzemi\'nski, W., et al. 2005,
in AIP Conf. Proc., 752, Stellar Astrophysics with the World's Largest Telescopes: First International
Workshop on Stellar Astrophysics with the World's Largest Telescopes, ed. J. Miko{\l}ajewska \& A. Olech (Melville, NY: AIP), 70
\bibitem[Koz{\l}owski et al.(2007)]{2007ApJ...671..420K} Koz{\l}owski, S., Wo\'zniak, P.~R., Mao, S., \& Wood, A. 2007, \apj, 671, 420
\bibitem[Minniti et al.(2010)]{2010NewA...15..433M} Minniti D., Lucas, P.~W., Emerson, J. et al. 2010, New Astron., 15, 433
\bibitem[Monaco et al.(2004)]{2004MNRAS.349.1278M} Monaco, L., Pancino, E., Ferraro, F.~R., \& Bellazzini, M. 2004, \mnras, 349, 1278
\bibitem[Paczy\'nski(1994)]{1994AcA....44..235P} Paczy\'nski, B. 1994, Acta Astron., 44, 235
\bibitem[Pietrinferni et al.(2006)]{2006ApJ...642..797P} Pietrinferni, A., Cassisi, S., Salaris, M., \& Castelli, F. 2006, \apj, 642, 797
\bibitem[Pietrukowicz et al.(2005)]{2005AcA....55..261P} Pietrukowicz. P., Kaluzny, J., Thompson, I.~B., et al. 2005, Acta Astron., 55, 261
\bibitem[Richter et al.(1999)]{1999A&A...350..476R} Richter, P., Hilker, M., \& Richtler, T. 1999, \aap, 350, 476
\bibitem[Rieke \& Lebofsky(1985)]{1985ApJ...288..618R} Rieke, G. H., \& Lebofsky, M.~J. 1985, \apj, 288, 618
\bibitem[Sahu et al.(2001)]{2001Natur.411.1022S} Sahu, K.~C., Casertano, S., Livio, M., et al. 2001, Nature, 411, 1022
\bibitem[Sahu et al.(2002)]{2002ApJ...565L..21S} Sahu, K.~C., Anderson, J., \& King, I.~R. 2002, \apj, 565, L21
\bibitem[Schlegel et al.(1998)]{1998ApJ...500..525S} Schlegel, D.~J., Finkbeiner, D.~P., \& Davis M. 1998, \apj, 500, 525
\bibitem[Stetson(1987)]{1987PASP...99..191S} Stetson, P.~B., 1987, \pasp, 99, 191
\bibitem[Udalski et al.(2000)]{2000AcA....50....1U} Udalski, A., \.Zebru\'n, K., Szyma\'nski, M., et al. 2000, Acta Astron., 50, 1

\end{thebibliography}
\end{document}